\documentclass[pra,amsmath,amssymb,twocolumn]{revtex4}
\usepackage{mathrsfs}
\usepackage{amssymb}

\input epsf.tex
\usepackage{graphicx}
\usepackage{amsthm}

\begin{document}

\title{Security of practical phase-coding quantum key distribution}

\author{Hong-Wei Li$^1$ $^2$, Zhen-Qiang Yin$^1$, Zheng-Fu Han$^1$*, Wan-Su Bao$^2$, Guang-Can Guo$^1$}

 \affiliation
 {$^1$ Key Laboratory of Quantum Information,University of Science and Technology of China,Hefei, 230026,
 China\\$^2$ Electronic Technology Institute, Information Engineer
University, Zhengzhou, 450004,
 China}

 \date{\today}
\begin{abstract}
Security proof of practical quantum key distribution (QKD) has
attracted a lot of attentions in recent years. Most of real-life QKD
implementations are based on phase-coding BB84 protocol, which
usually uses Unbalanced Mach-Zehnder Interferometer (UMZI) as the
information coder and decoder. However, the long arm and short arm
of UMZI will introduce different loss in practical experimental
realizations, the state emitted by Alice's side is nolonger standard
BB84 states. In this paper, we will give a security analysis in this
situation. Counterintuitively, active compensation for this
different loss will only lower the secret key bit rate.
\end{abstract}
\maketitle

\section{ Introduction}\label{Introduction}
Quantum key distribution \cite{review BB84} \cite{review Sc1} is the
art of allowing two distant parties Alice and Bob to remotely
establish a secret key combining with an authenticated classical
channel and a quantum channel. The unconditional security of quantum
key distribution bases on fundamental laws of quantum mechanics. The
unconditional security of QKD protocol with perfect experimental
setup (the source is perfect single photon source and so on) has
been respectively given by Lo, Chau \cite{review Lo}, Shor, Preskill
\cite{review Sh},
 Renner \cite{review Re} et al. Furthermore, Gottesman, Lo,
Lukenhaus and Preskill (GLLP) \cite{review GLLP} have given an
analysis of security of QKD bases on practical source. In their
security analysis, the final secret key bit can be generated if we
can estimate the lower bound of the secret key bits generated by the
single-photon pulse. More recently, Scarani \cite{review Sc2} has
analyzed security of QKD with finite resources. Meanwhile, the
experimental realization about QKD also has a rapid progress in
recent years \cite{decoy experiment1}\cite{decoy
experiment2}\cite{decoy experiment3}\cite{decoy
experiment4}\cite{decoy experiment5}\cite{decoy
experiment6}\cite{decoy experiment7}.

 In practical quantum key distribution realizations,
UMZI\cite{review Be}\cite{review Mo}\cite{review Han} method is
commonly used. However, the Phase Modulator (PM) in the
interferometer is not perfect, which means it will introduce much
more loss than the arm has no PM. As a result, in this case photon
states emitted by Alice's side is not the standard BB84 state (we
call it unbalanced states in the next section), which means security
of quantum key distribution based on this states can not be
satisfied with GLLP formula. Of course, one can give a simple
security proof when the loss of the PM is considered as an operation
controlled by Eve and then GLLP can be used to calculate the final
secret key rate. However, the secret key rate in this case is not
optimal.

 To give an optimal security proof of QKD with unbalanced BB84
states, we propose that the real-life source can be replaced by a
virtual source without lowering security of the protocol, then the
final secret key rate can be improved obviously.

\section {QKD with Unbalanced Mach-Zehnder Interferometer}\label{QKD with Unbalanced Mach-Zehnder Interferometer}

In this section, we will introduce the general QKD setup with UMZI
method, a schematic of this QKD setup can be illustrated as in Fig.
1.
\begin{figure}[!h]\center
\resizebox{8cm}{!}{
\includegraphics{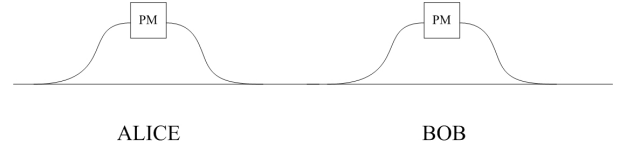}}
\caption{ QKD setup with Unbalanced Mach-Zehnder Interferometer.}
\end{figure}

In Alice's half-interferometer, there is a Phase Modulator (PM) in
the long arm. Correspondingly, the weak coherent state entering the
quantum channel can be divided into coherent state from the short
arm and from the long arm respectively, both of them can be given as
the following,

\begin{equation}
\begin{array}{lll}
|\alpha\rangle_s|\beta\rangle_l
\end{array}
\end{equation}

\begin{equation}
\begin{array}{lll}
|\alpha\rangle_s&=e^{-\frac{|\alpha|^2}{2}}\sum\limits_{n=0}^{\infty}\frac{\alpha^n}{\sqrt{n!}}|n\rangle_s\\
|\beta\rangle_l&=e^{-\frac{|\beta|^2}{2}}\sum\limits_{n=0}^{\infty}\frac{\beta^n}{\sqrt{n!}}|n\rangle_l\\
\end{array}
\end{equation}

\begin{equation}
\begin{array}{lll}
\alpha&=\sqrt{\mu}e^{i\theta}\\
\beta&=\sqrt{\nu}e^{i(\theta+\varphi)}
\end{array}
\end{equation}
where, $|\alpha\rangle_s$ is the weak coherent states in the short
arm, $|\beta\rangle_l$ is the weak coherent states in the long arm
after the PM, $|n\rangle_s$ is the $n$ photon state in the short
arm, $|n\rangle_l$ is the $n$ photon state in the long arm after the
PM, $\mu$ denotes the mean photon number of the short arm, $\nu$
denotes the mean photon number of the long arm, $e^{i\varphi}$ is
the phase modulated by the PM in the long arm. $e^{i\theta}$ is
selected uniformly at random because of Eve has no prior knowledge
of the phase. Combining with Lo and Preskill's method of phase
randomization in Ref. \cite{review LoPr}, the density matrix of the
state emitted by Alice is

\begin{equation}                     \label{munu1}
\begin{array}{lll}
\rho&=\int_0^{2\pi}\frac{d\theta}{2\pi}|\alpha\rangle_s|\beta\rangle_{ll}\langle\beta|_s\langle\alpha|\\
&=e^{-(\mu+\nu)}\sum\limits_{n=0}^{\infty}\frac{(\mu+\nu)^n}{n!}\Phi_n
\end{array}
\end{equation}

\begin{equation}                      \label{munu2}
\begin{array}{lll}
\Phi_n&=A_n|0\rangle_s|0\rangle_{ll}\langle0|_s\langle0|A_n^{+}\\
A_n&=(\frac{\sqrt{\mu}}{\sqrt{\mu+\nu}}a_s^{+}+e^{i\varphi}\frac{\sqrt{\nu}}{\sqrt{\mu+\nu}}a_l^{+})^n
\end{array}
\end{equation}
where, $\mu+\nu$ is the mean photon number in the quantum channel
emitted by Alice, $a^{+}$ is the creation operator.

We have given the practical state as in equations
(\ref{munu1},\ref{munu2}), thus the practical single photon state
after the PM is (the phase of the photon from the long arm is
randomly modulated by $0$, $0.5\pi$, $\pi$, $1.5\pi$)

\begin{equation}                                    \label{practical states}
\begin{array}{lll}
\frac{1}{\sqrt{\mu+\nu}} (\sqrt{\mu}|1\rangle_s+\sqrt{\nu}|1\rangle_l)\\
\frac{1}{\sqrt{\mu+\nu}}(\sqrt{\mu}|1\rangle_s+i\sqrt{\nu}|1\rangle_l)\\
\frac{1}{\sqrt{\mu+\nu}}(\sqrt{\mu}|1\rangle_s-\sqrt{\nu}|1\rangle_l)\\
\frac{1}{\sqrt{\mu+\nu}} (\sqrt{\mu}|1\rangle_s-i\sqrt{\nu}|1\rangle_l)\\
\end{array}
\end{equation}

The long arm and short arm have the same loss in the ideal case,
which means PM in Alice's side is perfect, thus we can get
$\mu=\nu$. However, the long arm will introduce much more loss than
the short arm in practical side, which means $\mu>\nu$ in practice,
the state emitted by Alice's side becomes unbalanced
correspondingly.

For simplicity, the practical PM can be replaced by a perfect PM
plus an unbalanced attenuator, the unbalanced attenuator only
attenuate the photon from the long arm, while the photon from the
short arm can be past without any attenuation. Since the single
photon state in this case is not the perfect BB84 state for the
eavesdropper Eve, formula for the final secret key rate (GLLP) can
not be satisfied in this case. To solve this problem, we will
analyze the final secret key rate with practical UMZI method. For
improving the final secret key rate, an unitary transformation will
be proposed in the next section.

\section {Security of QKD with Unbalanced Mach-Zehnder Interferometer}\label{Security of QKD with Unbalanced Mach-Zehnder Interferometer}

In this section, we will first give a very simple security analysis
of QKD with practical Unbalanced Mach-Zehnder Interferometer. As
mentioned in section \ref{QKD with Unbalanced Mach-Zehnder
Interferometer}, the half-interferometer in Alice's side can be seen
as an unbalanced attenuator in the quantum channel. A simple
security proof in this case can be illustrated as the following.

 We can simply assume the unbalanced attenuator is not controlled by Alice, which is part
of the quantum channel controlled by Eve, then the quantum state
emitted by Alice is standard BB84 states, and the final secret key
rate can be calculated. Based on this assumption, combining with
security analysis of the ideal decoy method QKD protocol \cite{decoy
theory1}\cite{decoy theory2}\cite{decoy theory3}\cite{decoy
theory31}\cite{decoy theory4}\cite{decoy theory5}\cite{decoy
theory6}, the upper bound of secret key bit rate generated by
standard single-photon BB84 states can be given by

\begin{equation}          \label{practicalkey}
\begin{array}{lll}
P_1{10}^{\frac{-{\alpha}l}{10}}P_AP_B=e^{-2\mu}2\mu
10^{\frac{-{\alpha}l}{10}}\frac{\nu}{2\mu}
\end{array}
\end{equation}
where, $P_1=e^{-2\mu}2\mu$ is the probability distribution of the
single photon state, $\alpha$ is the loss efficiency in the quantum
channel, $l$ is the fiber length, $P_A=\frac{\mu+\nu}{2\mu}$ is the
pass efficiency in Alice's side, $P_B=\frac{\nu}{\mu+\nu}$ is the
pass efficiency in Bob's side. Therefore, according to secret key
rate formula GLLP, the final secret key bit rate is just the key
rate with ideal PM but is lowered by the attenuation constant
$\nu/\mu$.

 However, the final secret key rate is low in this case. For improving the final secret
key rate, an unitary transformation will be proposed as in Fig. 2.

\begin{figure}[!h]\center
\resizebox{8cm}{!}{
\includegraphics{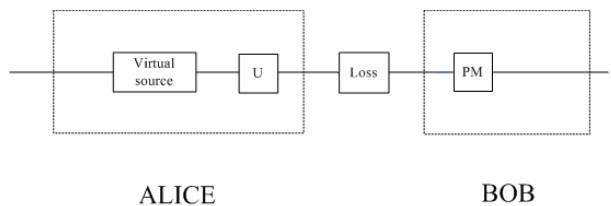}}
\caption{ UMZI method QKD with an assumption unitary transformation
and virtual source.}
\end{figure}

We will prove the practical state, as in equation (\ref{practical
states}), is as security as standard BB84 states by considering the
practical state is equal to the standard BB84 state combining with
an unitary transformation.

The unitary transformation in our paper is an virtual setup, which
does not need to be implemented in practical QKD experimental
realization, the detailed illustration of the unitary transformation
is given as the following,

\begin{equation}
\begin{array}{lll}
U|1\rangle_l|0\rangle_A&=\frac{\sqrt{\nu}}{\sqrt{\mu}}|1\rangle_l|0\rangle_A+\frac{\sqrt{\mu-\nu}}{\sqrt{\mu}}|0\rangle_l|1\rangle_A\\
U|n\rangle_l|0\rangle_A&=|n\rangle_l|0\rangle_A \quad
n\neq 1                                \\
 U|m\rangle_s|0\rangle_A&=|m\rangle_s|0\rangle_A
\quad m=0,1,\cdot \cdot \cdot
\end{array}
\end{equation}
where, $|0\rangle_A$, $|1\rangle_A$ are mutually orthogonal states
in Alice's system, which are unknown to Alice. The pass efficiency
of the single photon states is $P_{suc}=\frac{\mu+\nu}{2\mu}$.  The
practical photon state in Alice's side can be seen as the standard
BB84 state emitted by a virtual source combining with the unitary
transformation. Then security of practical QKD setup is equal to
security of QKD with the virtual source. Combining with the method
given in Ref. \cite{review GLLP}, we can prove the unconditional
security of new source QKD in the following.

As mentioned in section \ref{QKD with Unbalanced Mach-Zehnder
Interferometer}, the probability that single photon unbalanced
states emitted by Alice's side is $e^{-(\mu+\nu)}(\mu+\nu)$.
Combining with the unitary transformation, the probability
distribution of the single photon state of the virtual source
emitted by Alice's side is

\begin{equation}                      \label{practical1}
\begin{array}{lll}
\tilde{p_1}=\frac{e^{-(\mu+\nu)}(\mu+\nu)}{P_{suc}}\\
\end{array}
\end{equation}

 Similarly, the probability distribution of the
vacuum state and multi-photon state of the virtual source can be
given by
\begin{equation}
\begin{array}{lll}           \label{practical2}
\tilde{p_0}=e^{-(\mu+\nu)}-e^{-(\mu+\nu)}(\frac{(\mu+\nu)}{P_{suc}}-(\mu+\nu))\\
\tilde{p_n}=e^{-(\mu+\nu)}\frac{(\mu+\nu)^n}{n!}  \quad n\geqslant2
\end{array}
\end{equation}

 Obviously, the real-life setup of Alice can be replaced by the virtual
source and the basis-independent unitary transformation
equivalently. According to GLLP, the basis-independent unitary
transformation cannot lower the security of the QKD with the virtual
source, then we can calculate the upper bound of the secret key bit
rate, that is
\begin{equation}                       \label{ourkey}
\begin{array}{lll}
\tilde{p_1}p_{suc}10^{\frac{-{\alpha}l}{10}}P_B=e^{\mu+\nu}(\mu+\nu)10^{\frac{-{\alpha}l}{10}}\frac{\nu}{\mu+\nu}
\end{array}
\end{equation}

Comparing with equation (\ref{practicalkey}), upper bound of the
secret key bit rate can be improved $e^{\mu-\nu}$ times with our
security analysis.
 From equation (\ref{ourkey}), we can see the real-life unbalanced single photon states
can be generated from standard single photon states safely.
Therefore, security of the real-life unbalanced single photon states
is the same as standard single photon states, and GLLP formula is
satisfied in this situation. Combining with our security analysis,
one click of one real-life unbalanced single photon states can bring
one key bit safely.

One may argue that the problem can be taken away by adding the same
PM on the short arm, which can be controlled by Alice. However, the
same PM should be added in Bob's short arm for lowing the bit error
rate in this case, thus the secret key rate is

\begin{equation}          \label{practicalkey2}
\begin{array}{lll}
P_1{10}^{\frac{-{\alpha}l}{10}}P_A^{'}P_B^{'}=e^{-2\mu}2\mu
10^{\frac{-{\alpha}l}{10}}\frac{\nu}{2\mu}
\end{array}
\end{equation}
where $P_A^{'}=1$ is the pass efficiency in Alice's side,
$P_B^{'}=\frac{\nu}{2\mu}$ is the pass efficiency in Bob's side.
Obviously, equation (\ref{practicalkey2}) is the same as equation
(\ref{practicalkey}), thus the secret key rate is lower comparing
with our situation. On the other hand, all the unbalanced
multi-photons emitted by Alice's side cannot carry any secret key
bit, thus we can conclude that our key bit rate is optimal.

\section {simulation}  \label{simulation}

Similar to the method in Ref. \cite{decoy theory4}, we will give the
simulation result of the lower bound of the final secret key rate by
considering equations (\ref{practical1}) and (\ref{practical2}) in
this section. The mean photon number of $\mu$ is $0.4$ and the mean
photon number of $\nu$ is $0.067$ in the simulation. Combining with
GYS \cite{review GYS} parameters, the simulation result can be shown
in Fig. 3.

\begin{figure}[!h]\center
\resizebox{9cm}{!}{
\includegraphics{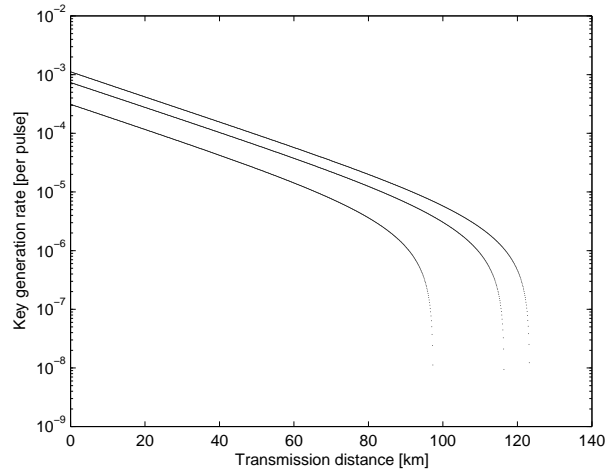}}
\caption{ Key generation rate for different transmission distance,
the upper line is the ideal case, where the PM is perfect, the
middle line is the calculation result with our method, the low line
is the calculation result without our method.}
\end{figure}

 From the simulation result,
we can see that the longest transmission distance can be improved
obviously with our security analysis.

\section {conclusions}\label{conclusions}
We have analyzed security of UMZI method QKD with practical phase
modulator by introducing a virtual source and an unitary
transformation in this paper. Correspondingly, the optimal key bit
rate has been given with our security analysis. From the simulation
result, we can see that our method can improve the final secret key
rate obviously. Quite interestingly, if the different loss was
compensated actively, the final secret key rate will be lower
comparing with no compensation.

\section {acknowledgements}

This work was supported by National Fundamental Research Program of
China (2006CB921900), National Natural Science Foundation of China
(60537020, 60621064) and the Innovation Funds of Chinese Academy of
Sciences. $^*$To whom correspondence should be addressed, Email:
zfhan@ustc.edu.cn.

\end{document}